\documentclass{aa} 

\usepackage{twoopt}
\usepackage{graphicx}
\usepackage{wrapfig}
\usepackage{float}
\usepackage{txfonts}
\usepackage{multirow}
\usepackage{tablefootnote}
\usepackage{amsmath}
\usepackage{siunitx}
\usepackage{indentfirst}
\usepackage{subfig}
\usepackage{blindtext}
\usepackage[dvipsnames]{xcolor}
\definecolor{darkgoldenrod}{rgb}{0.72, 0.53, 0.04}
\usepackage{appendix}
\usepackage{longtable}

\usepackage[colorlinks = true,
            linkcolor = RoyalBlue,
            urlcolor = darkgoldenrod,
            citecolor = RoyalBlue,
            anchorcolor = RoyalBlue, 
            draft = false]{hyperref}

\newcommand{\comment}[1]{}

\begin{document} 
   \title{Milky Way Near Twins (MWNeTs). I. A Hierarchical Framework for Identifying Candidate Evolutionary Counterparts of the Milky Way}

  \author{
I.B. Vavilova \inst{1} \thanks{Corresponding author} \and
A.M. Dmytrenko \inst{1} \email{admytrenko@mao.kiev.ua}\and
D.V. Dobrycheva \inst{1} \email{daria@mao.kiev.ua}\and
P.N. Fedorov \inst{2} \email{pnfedorov@gmail.com}\and
I.O. Izviekova \inst{1,3} \email{izviekova@gmail.com}\and
V.P. Khramtsov \inst{2} \email{vld.khramtsov@gmail.com}\and
O.V. Kompaniiets \inst{1} \email{kompanietsolena@gmail.com}\and
O.N. Kukhar \inst{1} \email{kukhar@mao.kiev.ua}\and
O.S. Pastoven \inst{1} \email{opastoven@mao.kiev.ua}\and
O. Sergijenko \inst{1} \email{olga.sergijenko.astro@gmail.com}\and
A.A. Vasylenko \inst{1} \email{vasylenko\_a@mao.kiev.ua}
}

\institute{
\inst{1} Main Astronomical Observatory of the NAS of Ukraine, 27, Akademik Zabolotnyi St., Kyiv 03143, Ukraine\\
\inst{2} Institute of Astronomy, V.N. Karazin Kharkiv National University, 35, Sumska St., Kharkiv, 61022, Ukraine \\
\inst{3} International Centre for Astronomical, Medical, and Ecological Research, National Academy of Sciences of Ukraine, 27, Akademik Zabolotnyi St., Kyiv 03123, Ukraine\\
}
\titlerunning{MWNeTs. I. A Hierarchical Framework}
\authorrunning{Vavilova et al.}

   \date{Received 2026; Submited to Astronomy \& Astrophysics}

\abstract
{Searches for Milky Way analogues (MWAs) traditionally rely on a limited set of present-day global properties. However, galaxies with similar morphology, stellar mass, star-formation rate, luminosity, and bulge-to-total ratio may have experienced substantially different assembly, secular, nuclear, and environmental histories.}
   {We introduce \textit{Milky Way Near Twins} (MWNeTs) as galaxies resembling the Milky Way in their present-day properties and exhibiting observational signatures consistent with broadly similar evolutionary pathways. We aim to extend parameter-based searches towards evolutionary similarity.}
{We propose seven stages: local environment, isolation criterion, and cosmic-web context; morphological and structural constraints; a Milky-Way-compatible nuclear state; global spectrophotometric and dynamical constraints; panchromatic spectral-energy-distribution similarity; rotation-curve similarity; and evolutionary fingerprints and fossil tracers. Each stage adds physically motivated constraints, progressively refining the candidate population and strengthening its assessment.}
   {For the first time, we formulate MWNeTs as a physically motivated class and establish a hierarchy explicitly distinguishing present-day compatibility from evidence for evolutionary similarity. The first four stages identify candidates compatible with the Milky Way's current state, while Stages V--VII provide complementary constraints on their evolutionary interpretation. We introduce \textit{evolutionary memory} to describe observational records of physical processes operating on different timescales. Spectral energy distributions, rotation curves, chemo-dynamical signatures, globular-cluster systems, merger histories, circumgalactic-medium properties, and multiwavelength fossil tracers jointly define a galaxy's \textit{evolutionary fingerprint}. These diagnostics constrain, rather than uniquely reconstruct, evolutionary pathways. A companion paper presents the observational application to NGC~3521, NGC~2903, UGC~11487, and CGCG~047-026.}
{The framework is inherently iterative: the Milky Way provides the reference, while near twins offer external laboratories for investigating aspects of Galactic formation and evolution inaccessible from our internal viewpoint. It connects Galactic astronomy, extragalactic astronomy, and cosmology through a comparative approach to testing broadly similar evolutionary pathways.}
\keywords{Galaxy: evolution -- Galaxy: fundamental parameters -- Galaxy: structure -- galaxies: evolution -- galaxies: multiwavelength properties }
\maketitle

\section{Introduction}
\noindent 
The Milky Way is the only large disc galaxy in which individual stars, stellar populations, components of the interstellar medium, and the central supermassive black hole can be studied in unprecedented detail. At the same time, our location within the Galactic disc complicates the determination of many global properties routinely measured in external galaxies. This duality makes comparisons with Milky Way analogues (MWAs) a particularly powerful tool for placing the Galaxy into a broader evolutionary and cosmological context.

Since the pioneering works by \citet{Vandenbergh1960classification, deVaucouleurs1972mw, deVaucouleurs1978, Pence1980mw}, in which the Milky Way was first considered as an extragalactic object through comparisons with nearby spiral galaxies of similar morphology, luminosity, and structural properties, the concept of MWAs has evolved substantially. The advent of large photometric and spectroscopic surveys such as SDSS, GAMA, MaNGA, and SAGA enabled the systematic identification of galaxies resembling the Milky Way and transformed MWAs into one of the principal observational frameworks for studying galaxy properties \citep{Gallagher1984starformation, Kennicutt1994galaxies, Vandenbergh2000galaxies, Hammer2007, Licquia2015mw, Geha2017saga, Zhou2023}. MWAs provide a framework for investigating secular evolution, calibrating galaxy scaling relations, and linking Galactic and extragalactic astronomy. They also address one of the fundamental questions in Galactic astronomy: to what extent is the Milky Way a typical spiral galaxy, and which of its properties may be considered unusual in a cosmological context?

The definition of ''Milky Way analogue galaxies” has progressively evolved from a simple morphological concept into a multidimensional astrophysical concept encompassing photometric, stellar population, environmental, and other properties. For example, many MWAs were initially selected using optical luminosities, colours, surface brightnesses, and the Tully–Fisher relation. Galaxies with similar luminosities can have substantially different internal structures, merger histories, environments, and evolutionary pathways. Consequently, no single parameter is sufficient to characterise a genuine Milky Way counterpart, implying that different selection criteria can produce substantially different MWA samples.

As a result, MWAs are increasingly regarded as multidimensional populations defined by different combinations of Milky Way-like properties. SDSS-based studies mainly identified morpho-photometric analogues \citep{Hammer2007, Mutch2011, Gonzalez2013mw, Licquia2015mw}, whereas integral-field surveys such as MaNGA extended the concept toward spatially resolved spectroscopic analogues \citep{Bundy2015manga, FraserMcKelvie2019, Boardman2020b, Boardman2020a, Zhou2023, GarmaOehmichen2023bar}. The SAGA survey further broadened this approach by incorporating satellite populations and environmental properties, thereby treating each Milky Way analogue together with its surrounding satellite system \citep{Geha2017saga, Mao2021saga}.

Identifying MWAs depends strongly on the number and nature of the selection criteria applied simultaneously \citep{Boardman2020b}. Selection based on only a few global parameters generally yields relatively large samples, supporting the conclusion that the Milky Way is not an unusual galaxy \citep{Licquia2016, Fraser2019, Pilyugin2019, Pilyugin2023, Tuntipong2024}. However, adding independent constraints quickly reduces the number of acceptable Milky Way candidates. Simultaneously matching morphology, stellar mass, star-formation activity, structural parameters, kinematics, chemical abundances, and environmental indicators yields progressively smaller yet physically more realistic MWA samples.

This trend shows that reproducing the Milky Way across multiple independent observables is substantially harder than matching only a few global parameters. More fundamentally, it raises the question of whether present-day similarity alone is sufficient to identify the Milky Way's closest evolutionary counterparts.

In this paper, we argue that the classical concept of MWAs, based primarily on present-day global properties, is no longer sufficient to identify the closest extragalactic counterparts to our Galaxy. The next logical step is the transition toward the concept of \textit{Milky Way Near Twins} (MWNeTs). We introduce MWNeTs as a physically motivated class of galaxies that resemble the Milky Way in their present-day global properties and exhibit independent observational signatures consistent with broadly similar evolutionary pathways, including environmental, structural, dynamical, chemical, nuclear, and multiwavelength characteristics  \citep{Vavilova2024}. In this framework, evolutionary diagnostics become an essential component of the selection procedure, complementing the classical photometric, spectroscopic, and morphological criteria traditionally used in MWA searches. The proposed transition may be summarised schematically as
\[
\mathrm{MWA}
\;\xrightarrow{\rm evolutionary\ diagnostics}\;
\mathrm{MWNeT}.
\]
Representative Milky Way parameters are summarised in Table~\ref{tab:MW_properties}, while the corresponding selection ranges and evolutionary diagnostics are presented in Table~\ref{tab:MWNeTcriteria}.

Introducing the MWNeT class enables a more rigorous investigation of the Milky Way in the context of external galaxies. Our recent forward-modelling study of the Milky Way near-twin candidate NGC~3521, based on a panchromatic SED extending from the ultraviolet to decameter radio wavelengths, demonstrated that observations of a near-twin galaxy can provide constraints on Galactic properties that are difficult or impossible to infer from our internal viewpoint alone \citep{Kompaniiets2026}. Such comparisons offer a unique opportunity not only to reconstruct the integrated appearance of the Milky Way as an external galaxy but also to identify evolutionary signatures and episodes of its history that remain hidden or only indirectly accessible from observations within the Galaxy itself. Integrated-light IFU spectroscopy provides an important observational bridge for testing the MWNeT concept, both by analysing the metallicity of MWNeTs \citep{Pilyugin2019, Xu2025} and the properties of the Milky Way as a distant galaxy \citep{Lian2023, Khoperskov2026}. Moreover, the multiwavelength diagnostics help uncover evolutionary episodes of the Milky Way that have left only subtle or indirect observational signatures.

The proposed MWNeT methodology follows a hierarchical strategy in which each successive stage adds a physically motivated constraint, progressively reducing the candidate population while strengthening its similarity to the Milky Way. Evolutionary diagnostics therefore constitute an essential component of the selection process. Rather than merely extending traditional parameter-based searches with additional observables, the framework reformulates the problem in terms of evolutionary similarity between external galaxies and the Milky Way. Unlike studies that apply additional criteria individually, we combine these diagnostics within a unified hierarchy.

In this first paper of the series, we establish the conceptual and methodological basis for identifying MWNeTs by combining present-day global properties with independent evolutionary diagnostics. This approach provides a systematic transition from broad MWA samples to identifying the closest candidate evolutionary counterparts of the Milky Way.

The paper is organised as follows. Section~2 presents the seven-stage MWNeT methodology and the physical motivation and observational basis of its diagnostics. Section~3 discusses the applicability, uncertainties, and limitations of the framework. Section~4 summarises the main conclusions. Appendix~A compiles the adopted reference properties of the Milky Way. The observational application to NGC~3521, UGC~11487, NGC~2903, and CGCG~047-026 is presented in a companion paper (Paper~II).

\begin{table*}
\caption{Hierarchical methodology for Milky Way Near Twins (MWNeTs) identification and evolutionary characterisation. Stages I--IV.}
\label{tab:MWNeTcriteria}
\centering
\begin{tabular}{p{0.36\textwidth}p{0.56\textwidth}}
\hline
Selection stage & Typical criterion/parameter \\
\hline
\multicolumn{2}{c}{\textbf{Stage I. Local Environment, isolation criteria, and Cosmic-Web Context ($\mathrm{MWA}\cap\mathrm{Environment}$)}}\\
\hline
Isolation criterion & No ongoing major interaction with a comparable-mass galaxy; loose-group membership and interactions with low-mass satellites are allowed (analogous to those between the Milky Way and the Magellanic Clouds). \\
Local environment & Loose-group or similarly low-density environment broadly comparable to that of the Local Group \\
Cosmic-web location &Low-density environment; void or filament; absence of rich-cluster or strongly interacting environment\\
Major satellite system & Possible presence of one or two massive satellites analogous to the LMC/SMC \\
\hline
\multicolumn{2}{c}{\textbf{Stage II. Morphological and Structural Constraints ($\mathrm{MWA}\cap\mathrm{Environment}\cap\mathrm{Barred}$)}}\\
\hline
Morphological type & SABbc--SBbc \\
Barred spiral structure & Barred or weakly barred spiral galaxies with grand-design / multi-arm structure \\
Disc scale length & $R_{\rm d}\sim2$--$5\,{\rm kpc}$ \\
Bulge-to-total ratio & $B/T\sim0.1$--$0.3$ \\
Bar structure & Long-lived stellar bar associated with a boxy/peanut-shaped bulge\\
Bar semi-major axis & $R_{\rm bar}\sim4$--$5.5\,{\rm kpc}$\\
Relative bar length & $R_{\rm bar}/R_{\rm disc}\sim0.18$--$0.35$ \\
Corotation-to-bar ratio & $\mathcal{R}=R_{\rm CR}/R_{\rm bar}\approx1.2$ for the Milky Way; values $1.0\lesssim\mathcal{R}\lesssim1.4$ correspond to fast bars \\
\hline
\multicolumn{2}{c}{\textbf{Stage III. Nuclear-State Constraints ($\mathrm{MWA}\cap\mathrm{Environment}\cap\mathrm{Barred} \cap\mathrm{Nuclear}$)}}\\
\hline
Supermassive black-hole mass & $M_{\rm SMBH} \sim 10^{6}$--$10^{7}\,M_\odot$\\
Nuclear activity & Quiescent nucleus, dominated by circumnuclear star formation,
classified as composite or LINER-like, or host only a weak, radiatively inefficient AGN; weak LLAGN; very low-accretion SMBH system \\
\hline
\multicolumn{2}{c}{\textbf{Stage IV. Global Spectrophotometric and Dynamical Constraints.  
(${\rm MWA}\cap{\rm Isolated}\cap{\rm Barred} \cap\mathcal{N}_{\rm MW}\cap{\rm Global}
\Rightarrow{\rm MWNeT}$)}}\\
\hline
Total luminosity & $L_*\sim(1$--$3)\times10^{10}\,L_\odot$ \\
Integrated luminosities & $L_{\rm IR} \sim 10^{9}$--$10^{10},L_\odot$;
$L_{\rm FIR} \sim 10^{9}$--$10^{10}\,L_\odot$;
$L_{\rm radio}(1.4{\rm GHz}) \sim 10^{21}{\rm W\,Hz^{-1}}$;
$L_X \sim 10^{39}$--$10^{40}\,{\rm erg\,s^{-1}}$ \\
Radio spectral index & $\alpha \sim 0.5$--$1.0$ \\
Optical/stellar disc size & $R_{25}\sim11$--$18\,{\rm kpc}$ or comparable optical-disc scale, depending on the adopted definition \\
Stellar mass & $M_*\sim(2$--$8)\times10^{10}\,M_\odot$ \\
Star-formation rate & ${\rm SFR} \sim 0.5$--$5\,M_\odot\,{\rm yr}^{-1}$ \\
Halo mass & $M_{\rm halo} \sim (0.8$--$1.8)\times10^{12}\,M_\odot$ \\
Characteristic circular velocity & $V_{\rm c}\approx220$--$245\,{\rm km\,s^{-1}}$ \\
Mean stellar / gas-phase metallicity & Broadly near-solar; strong radial and population-dependent variations are allowed \\
Gas fraction & $f_{\rm gas} \sim 0.05$--$0.2$ \\
Neutral hydrogen mass & $M_{\rm HI} \sim (3$--$10)\times10^{9}\,M_\odot$ \\
Molecular gas mass & $M_{\rm H_2} \sim (0.5$--$3)\times10^{9}\,M_\odot$ \\
\hline
\end{tabular}
\end{table*}

\begin{table*}
\noindent
Table~\ref{tab:MWNeTcriteria}: Hierarchical methodology for
Milky Way Near Twins (MWNeTs) identification and evolutionary
characterisation (continued). Stages V--VII.
\par\medskip
\centering
\begin{tabular}{p{0.36\textwidth}p{0.56\textwidth}}
\hline
Selection stage & Typical criterion/parameter \\
\hline
\multicolumn{2}{c}{\textbf{Stage V. Panchromatic SED Similarity}}\\
\hline
Integrated SED shape & Broad similarity to the integrated Milky Way SED from the UV to the radio, including a moderate UV contribution, a dominant optical/NIR stellar component, FIR dust emission, and no energetically dominant persistent AGN \\
Normalised SED comparison & Agreement between the normalised luminosity distributions of the candidate and the Milky Way across the common wavelength range \\
Local SED slopes & Similarity of the logarithmic SED slopes, quantified where possible using correlation coefficients, slope residuals, and wavelength-resolved deviations \\
Component-level differences & Identification of physically meaningful departures in the UV, dust-emission peak, radio continuum, or other wavelength regimes without requiring point-by-point identity \\
\hline
\multicolumn{2}{c}{\textbf{Stage VI. Rotation-Curve Similarity}}\\
\hline
Characteristic rotation amplitude & Circular or deprojected rotation velocity broadly compatible with the Milky Way velocity scale \\
Rotation-curve morphology & Similarity in the radial shape of the rotation curve, including the position of its maximum, approximately flat intermediate-radius behaviour, and a possible mild outer decline \\
Outer radial slope & Mean radial velocity gradient compared over a common physical radial interval, where sufficiently resolved kinematic data are available \\
Kinematic limitations & Explicit consideration of inclination, asymmetric drift, non-circular motions, warps, anomalous gas components, tracer dependence, and radial coverage \\
\hline
\multicolumn{2}{c}{\textbf{Stage VII. Evolutionary Fingerprints and Fossil Tracers}}\\
\hline
Merger and accretion history & Relatively quiescent recent evolution, with no securely established major merger during the last $\sim8$--$10\,{\rm Gyr}$; ancient accretion events and subsequent minor mergers or satellite interactions are allowed \\
Chemo-dynamical signatures & Radial abundance gradients, breaks or flattening in the chemical distribution, inner-disc and bar stellar populations, thick/thin-disc properties, and fossil signatures of past accretion \\
Resolved stellar populations & Age, metallicity, and kinematic substructures tracing secular evolution, disc growth, star-formation episodes, or accretion events \\
Globular-cluster system & A Milky-Way-compatible globular-cluster population, including possible in-situ and accreted subpopulations; approximately $150$--$180$ clusters may be used as a reference rather than a strict selection interval \\
Circumgalactic medium & Extended multiphase circumgalactic gas and evidence for gas accretion, recycling, or disc--halo exchange \\
Large-scale multiwavelength fossil tracers & Anomalous H\,{\sc i}, stellar streams, shells and low-surface-brightness envelopes, radio spurs or loops, synchrotron haloes, large-scale outflows, X-ray echoes, and relics of past nuclear activity or TDE-like events \\
\hline
\multicolumn{2}{p{0.92\textwidth}}{\footnotesize
\textbf{Interpretation of the hierarchical stages.}
Stage~I identifies galaxies residing in an environmental and cosmic-web context broadly compatible with that of the Milky Way. Stage~II evaluates morphological and structural similarity, including the presence and properties of a stellar bar. Stage~III tests whether the nuclear state is compatible with the weak, quiescent, intermittent, or radiatively inefficient regime of the Milky Way; it does not require a formal LLAGN classification, whereas a persistent radiatively efficient Seyfert or quasar nucleus is treated as incompatible. Stage~IV determines whether the candidate reproduces the present-day global physical properties of the Milky Way. Stages~V and VI provide independent quantitative tests based on the panchromatic SED and resolved rotation-curve morphology, respectively. Stage~VII examines evolutionary fingerprints preserved in chemical, stellar, gaseous, circumgalactic, and other fossil tracers. The integrated MWNeT assessment synthesises the evidence from Stages~I--VII and is not treated as an additional physical stage. In general, Table~\ref{tab:MW_properties} summarises the adopted Milky Way reference properties, whereas the broader intervals listed here define the proposed compatibility ranges for candidate galaxies.}\\
\hline
\end{tabular}
\end{table*}

\section{The hierarchical MWNeT framework}
The primary objective of this work is to shift the search for MWNeTs from the traditional multidimensional parameter space to an evolutionary-diagnostic framework, in which the Milky Way serves as the astrophysical reference system. In this framework, similarity in evolutionary history becomes more informative than similarity in present-day global properties alone. This conceptual transition extends the classical MWA approach and defines the MWNeT class through a hierarchy of physically motivated selection criteria.

Table~\ref{tab:MWNeTcriteria} presents the hierarchical methodology for MWNeT identification proposed in this work. Unlike classical MWA searches that match a limited set of global observables, each successive stage adds complementary physical constraints. Consequently, the candidate sample is progressively reduced, while the physical and evolutionary similarity of the remaining galaxies to the Milky Way increases.

The multidimensional nature of the problem can be illustrated by grouping the properties commonly considered in studies of Milky Way analogues into several broad categories:

\textit{Global structural properties}: morphological type, disc scale length, bulge-to-disc ratio, bar characteristics, stellar mass, and dark-matter halo mass;

\textit{Stellar-population indicators}: star-formation rate and specific star-formation rate, stellar age distribution, metallicity gradients, $\alpha$-element abundances, and thick/thin disc properties;

\textit{Interstellar-medium parameters}: atomic and molecular gas masses, gas fraction, dust properties, ISM phase structure, magnetic-field properties, synchrotron halos, and circumgalactic-medium characteristics;

\textit{Multiwavelength characteristics}: the spectral energy distribution (SED) from UV to radio and X-ray wavelengths, IR and radio luminosities, spectral indices, FIR--radio correlations, and other multiwavelength peculiarities;

\textit{Dynamical constraints}: rotation-curve morphology, angular momentum, velocity dispersion, merger history, and secular-evolution processes;

\textit{Nuclear properties}: nuclear spectral type, supermassive black-hole mass, accretion rate, and Eddington ratio;

\textit{Environmental properties}: local environment, satellite-system structure including Magellanic-Cloud-like companions, Local-Group-like volume density, and cosmic-web location.

The large number of potentially relevant observables makes identifying genuine MWNeTs a highly multidimensional problem. Moreover, homogeneous measurements of many of these properties are unavailable for large galaxy samples, and not all diagnostics can be applied simultaneously. Therefore, simply increasing the dimensionality of the conventional parameter space does not necessarily provide a physically meaningful route to identifying the closest evolutionary MW counterparts. This motivates the hierarchical methodology adopted here, in which candidate galaxies are progressively filtered with independent physical constraints, while the most informative evolutionary diagnostics are applied to increasingly restricted samples.

The resulting methodology, summarised in Table~\ref{tab:MWNeTcriteria}, consists of seven successive stages: (i) local environment, isolation, and cosmic-web context; (ii) morphological and structural constraints; (iii) a Milky-Way-compatible nuclear state; (iv) global spectrophotometric and dynamical constraints; (v) panchromatic SED similarity; (vi) rotation-curve similarity; and (vii) evolutionary fingerprints and fossil tracers. The first four stages are mandatory for initial MWNeT candidate selection and assess whether a galaxy's present-day environmental, structural, nuclear, and global properties are compatible with those of the Milky Way. Stages~V--VII test and refine their evolutionary interpretation through complementary energetic, dynamical, chemo-dynamical, and fossil diagnostics. Compatibility at Stage~IV therefore establishes candidate status rather than demonstrating evolutionary similarity. The integrated assessment combines the evidence across all seven stages, explicitly accounting for observational uncertainties, incomplete data, and diagnostic degeneracies. The hierarchy thus extends the search beyond minimising offsets in a multidimensional parameter space towards testing whether similar present-day galaxies exhibit observational signatures consistent with broadly comparable evolutionary pathways.

\subsection{Isolation Constraints and Environment. The First Step Beyond the Classical MWA Concept}

Isolation and morphology probe different aspects of the Milky Way's similarity. Isolation assesses whether a galaxy has evolved in an environment comparable to the Milky Way's. Thus, isolation serves primarily as a criterion of evolutionary context, reflecting the role of environment in shaping galaxy formation and evolution. Morphology addresses whether a galaxy has developed a structural configuration comparable to that of the Milky Way. Morphological properties, therefore, represent a criterion of structural similarity, characterising the present-day appearance and internal organisation of the galaxy.

Isolation criteria are extremely important -- and often underestimated -- both for the search for MWAs and for defining truly Milky Way-like systems. 

\textbf{Large-scale environment.} We consider the location of the Milky Way in the cosmic web as one of the key factors governing its evolutionary state and adopt the hypothesis that the global properties of the Milky Way result from an evolution without major mergers during the last $\sim$10 Gyr \citep{Hammer2007, Bland2016, Helmi2020}. High-resolution N-body simulations of the last major merger event \citep{Naidu2021} constrained the orbital parameters of the Gaia–Enceladus/Sausage (GES) merger and revealed characteristic density-profile breaks at $\sim$15--18 and $\sim$30 kpc. These density-profile breaks may reflect not only the Galaxy's internal structure, including star formation history, gas accretion, and chemical abundance asymmetries, but also rotation-curve behaviour and dark-matter distribution, especially in the outer regions of the Galactic disc and halo. 

Current cosmographic reconstructions indicate that the Local Group lies within the Local Sheet, part of a nearby filamentary structure bordering the Local Void, rather than occupying the centre of a dense cluster environment \citep{Tully2008, McCall2014, Tully2019}. The gravitational interplay between the surrounding filaments, the expanding Local Void, and the nearby Virgo overdensity has likely influenced the Milky Way's long-term gas accretion, satellite population, angular momentum evolution, and secular development \citep{Tully2019, AragonCalvo2022}. The influence of the Local Void, adjacent to the Local Group \citep{Tully1987, Lindner1995, Mazurenko2024}, on the Milky Way's dynamical evolution remains an open question. A series of Cosmicflows studies showed that the Local Void may exert a substantial dynamical effect, contributing to the Local Group's peculiar velocity and driving the Milky Way away from this underdense region \citep{Tully2008, Courtois2017, Tully2019, Anand2019}. In this context, underdense regions can act as ``repellers'' in the large-scale velocity field, complementing the gravitational attraction of massive structures such as Virgo and the Shapley concentration \citep{Hoffman2017}. 

In general, the dynamical role of cosmic voids in shaping galaxy motions is increasingly recognised as comparable in importance to the attraction exerted by overdense structures. At the same time, the Milky Way's large-scale environment is shaped by the filamentary structure of the Local Sheet and neighbouring cosmic filaments, which may regulate gas accretion, angular-momentum acquisition, and preferred satellite infall directions \citep{McCall2014, Libeskind2015, Kraljic2020}. In the context of $\Lambda$CDM cosmology, the Milky Way's relatively quiescent merger history may represent a non-typical evolutionary pathway for a galaxy of comparable mass \citep{Hammer2007, BoylanKolchin2013, Fattahi2016}.

\textbf{Local Group-like environment.} Even minor companions may significantly affect the evolution of disc galaxies through tidal perturbations, gas transfer, dark-matter wake formation, and triggering of star-formation episodes \citep{Besla2012, Laporte2018, GaravitoCamargo2019}. This is one of the most interesting issues in MWA and MWNeTs research: is the Milky Way ``typical" because it is relatively isolated, or unusual because of the specific configuration of the Local Group? The proximity of M31, the presence of the Magellanic Clouds, the Gaia--Enceladus merger history, and the rich satellite population may together represent an atypical configuration for a galaxy of Milky Way mass. Consequently, overly strict isolation criteria may accidentally exclude the most realistic analogues. 

In addition, galaxies with similar stellar or halo masses may nevertheless follow substantially different evolutionary pathways due to assembly bias and environmental dependence within the cosmic web \citep{Gao2005, Wechsler2018}. The orientation of galactic angular momentum relative to neighbouring filaments may further influence disc stability, gas accretion geometry, and anisotropic satellite infall \citep{Tempel2013, Dubois2014}. The relatively moderate present-day star-formation rate of the Milky Way may therefore reflect its intermediate evolutionary state between strongly interacting systems and fully isolated quenched disc galaxies.

A strict photometric isolation criterion is often too simplistic. A galaxy that appears isolated at the present epoch may nevertheless have experienced significant mergers or long-term tidal interactions in the past. Conversely, mild group membership may be essential for reproducing the Milky Way's evolutionary pathway. The role of minor mergers and satellites, therefore, remains an important, unresolved issue \citep{Besla2007, Patel2020, Conroy2021}. 

In particular, the influence of the Magellanic Clouds as gas reservoirs is still debated. \cite{vandenBergh2006} suggested that the LMC and SMC may be interlopers from a remote part of the Local Group as opposed to true satellites of the Milky Way, implying that the Large Magellanic Cloud (LMC) may currently be on its first approach to the Milky Way. \cite{Font2021} and \cite{Jones2024} investigated the significance of satellite effects and analysed star-formation rates in galaxy systems similar to the MW-like satellite system using projected satellite--host distances. Using SAGA II, CFHT, and VLA data, \cite{Jones2024} identified several MW-like systems where satellites undergo ram-pressure stripping while interacting with the circumgalactic medium (CGM) of the host galaxy. 

In fact, the CGM may additionally preserve signatures of past accretion and interaction events over cosmological timescales. Therefore, similarities in stellar morphology or global photometric parameters alone may be insufficient for identifying true MWAs without considering CGM properties, gas flows, and baryon cycling processes \citep{Tumlinson2017, Putman2021}. 

A complete census of the MW satellite system remains elusive, as new dwarf galaxies continue to be discovered. Currently, more than 60 satellites are known, including the Magellanic Clouds, the Sagittarius Dwarf Spheroidal Galaxy, Fornax, Sculptor, Draco, Ursa Minor, Carina, Sextans, Leo I and Leo II, as well as numerous ultra-faint dwarf galaxies, including Antlia 2, one of the strangest known dwarf galaxies discovered recently by Gaia.
 
The boundary between a dwarf galaxy, a tidally disrupted dwarf, and a globular cluster is sometimes uncertain. For example, $\omega$ Centauri is often interpreted as the remnant core of an accreted dwarf galaxy, while Sagittarius and Boötes III are currently undergoing tidal disruption by the Milky Way. The Gaia mission has revolutionised the search for MW satellites, stellar streams, diffuse dwarf galaxies, and tidal remnants. Nevertheless, despite the growing number of detected satellites, the missing-satellite problem remains unresolved: $\Lambda$CDM cosmology predicts substantially more dwarf satellites than are currently observed around the Milky Way. Possible explanations include extremely low surface brightness, low stellar content, and observational incompleteness.

It was long believed that the Milky Way experienced an unusually quiet evolutionary history. However, modern observations reveal signatures of the Gaia--Sausage--Enceladus merger, ongoing Sagittarius dwarf interaction, and the massive infall of the LMC. In cosmological terms, this may instead indicate a dynamically complex merger history, non-trivial chemical evolution, and an unusually disc-dominated galaxy structure. Moreover, the presence of two relatively massive star-forming Magellanic Clouds may itself be somewhat uncommon among Milky Way-mass galaxies; at least, such systems are very rare in the IllustrisTNG simulations \citep{Haslbauer2024}. 

Among the most interesting systems are ``MW + LMC'' analogues, because massive satellites comparable to the LMC may also be relatively rare. Their presence strongly influences halo dynamics and may require identifying chemo-dynamical analogues rather than purely photometric MWAs. Such systems are particularly important for comparing metallicity distributions, thick-disc properties, stellar-halo substructure, dark matter signals, and accretion histories \citep{ForbesBridges2010, Salem2015, Vienn2026, Dillamore2026}. The increasing complexity of MWA selection suggested that studies may require machine-learning and latent-space approaches capable of comparing galaxies through multidimensional chemo-dynamical similarity metrics and simulations \citep{Brooks2026}.

The future MW--M31 interaction over the next $\sim5$ Gyr also remains uncertain \citep{Dubinski1996, vanDerMarel2012, Cautun2019}. Based on Gaia and HST data, \cite{Sawala2024} showed that the presence of the LMC and M33 can significantly affect the MW--M31 orbital evolution because the LMC orbit is nearly perpendicular to the MW--M31 orbital plane. Moreover, these authors found that existing uncertainties in the present kinematic and dynamical parameters of Local Group galaxies imply nearly a 50\% probability of a future MW--M31 merger within the next 10 Gyr. An additional role may be played not only by uncertainties in distance determinations \citep{Elyiv2020, DiValentino2025, H0DN}, but also by the MW--M31 orbital geometry \citep{Banik2022} and the position of the Local Void adjacent to the Local Group \citep{Tully1987, Lindner1995, Mazurenko2024}.

Consequently, the Milky Way itself may not represent a universal template for spiral galaxies, but rather a specific evolutionary realisation shaped by a rare combination of large-scale environment, merger history, Local Group dynamics, and long-term gas accretion conditions. 

\textbf{Brief conclusion for isolated MWNeTs search.} Thus, the Milky Way itself is not perfectly isolated. It belongs to the Local Group and has both a massive nearby companion, M31, and a complex local environment, including the Magellanic Clouds and numerous dwarf satellites. The Milky Way is therefore better described as \textit{moderately isolated in a loose-group environment}. Similarity in the cosmic-web location may represent a necessary evolutionary condition for MWNeTs.

Isolation criteria for MWNeTs should therefore account not only for projected galaxy density but also for massive companions, group membership, merger history, and the surrounding cosmic-web environment. Simple photometric isolation criteria may exclude galaxies that most closely reproduce the Milky Way's true evolutionary pathway. As the first step, we considered a sample of spiral barred galaxies in the Local Volume and estimated local environment parameter with 3D Voronoi tessellation \citep{Vavilova2021b} and k-NN method (k=5): $M_*^{\mathrm{neighbour}} < \frac{1}{2}M_*^{\mathrm{target}}$ within several hundred kpc or less than 1 Mpc to exclude strongly interacting galaxies \citep{Kompaniiets2025}.

Isolation is therefore fundamentally multidimensional and cannot be characterised by a single universally adopted quantitative parameter. Unlike stellar mass, SFR, or other global galaxy properties, catalogues and environmental studies define and estimate isolation differently. The AMIGA project commonly uses the local number-density estimator $\eta_k$ and the tidal-strength parameter $Q$ \citep{Verley2007, ArgudoFernandez2013}, while the CIG and 2MIG catalogues apply catalogue-specific isolation criteria based primarily on the angular sizes and projected separations of neighbouring galaxies \citep{Karachentseva1973, Karachentseva2010}. Other studies quantify the environment using tidal indices, projected galaxy densities, nearest-neighbour distances, or related environmental estimators \citep{Muldrew2012}. Therefore, the present MWNeT framework adopts no unique quantitative isolation criterion. Depending on the available observational data and the parent galaxy sample, isolation may be assessed using commonly employed environmental indicators or catalogue-specific criteria, provided these criteria identify galaxies whose evolution has not been dominated by strong external perturbations from massive neighbouring systems. 

In the MWNeT methodology, isolation should therefore be understood as a multidimensional evolutionary constraint that incorporates the tidal environment, group membership, satellite population, merger history, and position within the cosmic web. It represents the first constraint on whether a galaxy has evolved in an environmental context comparable to that of the Milky Way and therefore forms the foundation of the hierarchical MWNeT methodology. Once this condition is satisfied, the next step is to determine whether the galaxy's internal structure has evolved toward a Milky Way-like morphology.

\subsection{Morphological Constraints}
Following the environmental selection described above, the second stage of the proposed methodology restricts the remaining sample to barred or weakly barred intermediate-type spiral galaxies, approximately SABbc--SBbc, consistent with the currently accepted morphology of the Milky Way.

\textbf{The Galactic bar as a morphological and evolutionary diagnostic.} Particular attention should be paid to the Galactic bar, which may represent one of the most important indicators of structural similarity between candidate galaxies and the Milky Way. Although bar classification is commonly included in the morphological selection of MWAs, its evolutionary significance is often underestimated.

The Galactic bar is associated with a boxy/peanut-shaped bulge and plays a key role in the secular evolution of the Milky Way \citep{Wegg2015, Bland2016, Sanders2019}, shaping its about $t_{\rm bar}\sim 8\pm2~{\rm Gyr}$ \citep{Athanassoula2005, Athanassoula2013} (although estimates ranging from $\sim6$--$10~{\rm Gyr}$ have been reported in the literature \citep{Bland2016, Debattista2017}). Consequently, similarity in bar properties may provide valuable information not only about the present-day morphology of candidate galaxies but also about their evolutionary histories. The bar strongly influences the redistribution of angular momentum and drives radial gas flows toward the central regions of the Galaxy, star formation in the inner disc, and growth of the pseudobulge, potentially contributing to the fueling cycle of Sgr~A* \citep{Kormendy2004, Kruijssen2014, Sormani2015, Bland2016, Vera2016, Sormani2019}. 

Within the proposed MWNeT framework, we treat the presence of a bar as a necessary structural condition, reflecting its fundamental role in shaping the Milky Way's present-day structure. More detailed bar parameters, including the relative bar size, $R_{\rm bar}/R_{\rm disc}$, and the corotation-to-bar ratio, $\mathcal{R}=R_{\rm CR}/R_{\rm bar}$, provide additional constraints on the secular evolution and dynamical state of candidate galaxies. Thus, bar presence enters the morphological selection at Stage~II, whereas quantitative similarity in bar structure and dynamics strengthens the evolutionary classification of MWNeTs.

The detectability of a stellar bar depends strongly on viewing geometry. As an example, if the Milky Way were observed externally at an inclination comparable to that of NGC~3521 ($i\simeq73^\circ$), its bar could appear substantially foreshortened and partially obscured by the bulge, disc, and dust lanes. The effect would be particularly severe for an approximately end-on bar orientation. Consequently, the absence of an unambiguous photometric bar classification in a highly inclined galaxy cannot by itself be interpreted as evidence for the physical absence of a bar.

We therefore distinguish between a genuinely unbarred system and a galaxy for which bar identification is limited by projection. The latter is classified as \textsc{Partial} or \textsc{Unconstrained}, depending on whether independent photometric or kinematic bar signatures are present. A \textsc{Fail} classification is assigned only when sufficiently deep, inclination-corrected structural and kinematic data provide robust evidence against a bar.

\textbf{Relative bar size.} Beyond conventional Hubble classification, quantitative bar parameters may provide additional constraints on Milky Way similarity. The relative bar size, $R_{\rm bar}/R_{\rm disc}$, characterises the prominence of the bar with respect to the stellar disc and may reflect the degree of secular evolution. For the Milky Way, the bar semi-major axis is typically estimated as $R_{\rm bar}\simeq4.5$--$5.5~{\rm kpc}$. Depending on whether the optical disc radius or the full extent of the stellar disc is adopted, estimates of $R_{\rm disc}$ range from approximately $15$--$18~{\rm kpc}$ to $25$--$30~{\rm kpc}$, corresponding to $R_{\rm bar}/R_{\rm disc}\simeq0.18$--$0.35$ (Table~\ref{tab:MWNeTcriteria}). Because this ratio depends on the adopted definition of disc size, use it as a comparative structural diagnostic rather than a strict universal selection threshold. Although this parameter is rarely used in MWA searches, it may provide an additional constraint on the degree of secular evolution and the disc's dynamical state. 

A complementary dynamical diagnostic is the ratio between the corotation radius, $R_{\rm CR}$, and the bar length, $\mathcal{R} = \frac{R_{\rm CR}}{R_{\rm bar}}$. This dimensionless quantity is commonly used to distinguish between ''fast" and ''slow" bars and provides a direct constraint on bar dynamics and secular evolution. Bars with $1.0 \lesssim \mathcal{R} \lesssim 1.4$ are generally classified as fast bars \citep{Debattista2000, Aguerri2015}. For the Milky Way, $\mathcal{R} = 1.2$, which places the Galactic bar in the regime of fast rotators \citep{Portail2017, Shen2020}. Since the parameter $\mathcal{R}$ reflects the dynamical state of the bar and its interaction with the stellar disc and dark-matter halo, it may provide an additional criterion for identifying MWNeTs. 

 \textbf{Angular momentum.} The galactic bar regulates radial migration, gas inflow, central mass assembly, and the formation of box/peanut bulges by redistributing angular momentum among the stellar disc, bulge, gaseous component, and dark matter halo \citep{Gerhard2016}. These processes shape the Galaxy's structural and chemo-dynamical evolution over several gigayears. Recent theoretical and observational studies by \cite{Romeo2023} further emphasise that the specific angular momentum of galaxies is closely linked to galaxy morphology, bar formation, disc gravitational instability, and the efficiency with which baryons retain angular momentum during galaxy evolution. Since dark matter halo spin regulates the angular momentum retained by baryons and stars, it ultimately shapes galaxy morphology and secular evolution. These results identify angular momentum as both a dynamical and evolutionary diagnostic, further motivating its inclusion among the criteria defining MWNeTs.

\textbf{Brief conclusion.} The morphological stage of the MWNeT methodology therefore selects barred or weakly barred intermediate-type spirals, approximately SABbc--SBbc, as the closest structural counterparts of the Milky Way. Bar presence imposes a necessary morphological constraint, whereas relative bar size and the corotation-to-bar ratio provide additional information on the dynamical state and secular evolution of candidate galaxies. The combination of morphological classification with quantitative bar diagnostics therefore provides a physically motivated transition from structural similarity toward evolutionary similarity.

\subsection{Nuclear Activity and SMBH Constraints} 
The nuclear activity of the Milky Way is associated with its central supermassive black hole, Sagittarius~A$^\ast$, with $M_{\rm SMBH}\simeq4\times10^{6}\,M_{\odot}$ \citep{Ghez2008, Genzel2010, GravityCollaboration2019, Becerra-Vergara2021}. At the present epoch, Sgr~A$^\ast$ is characterised by an extremely low accretion rate, $\dot{M}\sim10^{-9}$--$10^{-7}\,M_{\odot}\,\mathrm{yr}^{-1}$, and an Eddington ratio $\lambda_{\rm Edd}\ll10^{-6}$ \citep{Baganoff2003, Ho2008, Yuan2014}. Although the Galactic nucleus is predominantly quiescent, rapid flares are occasionally observed across multiple wavelength ranges \citep{Genzel2010, Dodds2011}. Unlike classical luminous AGNs, the Milky Way does not host a prominent parsec-scale dusty torus. Instead, Sgr~A$^\ast$ is surrounded by a clumpy circumnuclear disc (CND) of molecular gas and dust extending over approximately $1$--$5$ pc \citep{Etxaluze2011, Lau2013, Tsuboi2018}. The Galactic nucleus may therefore be regarded as an extremely low-luminosity active nucleus or a nearly quiescent SMBH system. Low-level nuclear activity is relatively common among massive spiral galaxies in the nearby Universe \citep{Ho2008, Pulatova2015}.

Weak AGN activity may represent a key evolutionary constraint for MWNeT selection because it reflects the long-term coupling among SMBH growth, gas accretion, secular evolution, star formation, and feedback processes \citep{Kormendy2004, KormendyHo2013, HeckmanBest2014}. Many spiral galaxies with otherwise Milky Way-like morphology and stellar mass host substantially stronger nuclear activity, including Seyfert nuclei, LINER emission, compact radio cores, or jet activity. Therefore, requiring complete nuclear inactivity would be too restrictive, whereas allowing powerful AGN activity would select galaxies with significantly different evolutionary histories. Accordingly, Stage~III of the MWNeT methodology (Table \ref{tab:MWNeTcriteria}) excludes galaxies hosting persistently powerful AGNs, while retaining quiescent nuclei and weakly active SMBH systems whose present-day nuclear state is broadly compatible with that of the Milky Way.

However, the Milky Way was probably not always so quiescent. The Fermi bubbles and X-ray reflection features observed in molecular clouds near the Galactic Centre provide evidence for previous episodes of enhanced nuclear activity \citep{Su2010, Ponti2010, Clavel2013, Ponti2019}. Their origin remains debated and may involve nuclear star formation or past AGN-like outbursts from Sgr~A$^\ast$ \citep{Zubovas2012, Guo2012, Yang2012}. Some models suggest that the Galactic Centre may have temporarily reached luminosities of $L\sim10^{41}$--$10^{43}\,\mathrm{erg\,s^{-1}}$ during short-lived active phases. These observations show that the present-day nuclear state alone cannot recover the Milky Way's long-term activity history. For example, recent time-domain mid-IR studies of the isolated barred galaxy UGC~11487/WTP14adeqka \citep{Masterson2024} suggest that apparently quiescent nuclei may preserve observational signatures of recent obscured accretion events, reinforcing the view that present-day nuclear activity alone cannot characterise the long-term SMBH evolutionary history.

These observational indications suggest that nuclear activity in disc galaxies is episodic and may evolve over cosmological timescales \citep{Ho2008, Ponti2019}. Consequently, a genuine Milky Way Near Twin is not required to reproduce the exact present-day activity state of the Galactic nucleus. Rather, it should be consistent with a comparable long-term regime of SMBH growth, accretion duty cycle, and low-level feedback, reflecting a broadly similar evolutionary pathway.

So, the nuclear stage of the MWNeT hierarchy should not be formulated as a requirement that a candidate galaxy host, for example, a low-luminosity active galactic nucleus (LLAGN). The present-day nucleus of the Milky Way is radiatively quiescent, while observational evidence from the Galactic Centre and large-scale fossil structures indicates that it has experienced episodic activity in the past. Consequently, it should reflect the time-dependent nature of nuclear activity. This class includes galaxies whose nuclei are currently quiescent, dominated by circumnuclear star formation, classified as composite or LINER-like, or host only a weak LLAGN, a radiatively inefficient AGN. The assessment should combine, where available, optical emission-line diagnostics, nuclear X-ray and radio emission, mid-infrared colours or variability, the SMBH-mass estimate, and the Eddington ratio. Optical BPT or WHAN classifications alone are insufficient, because LINER-like line ratios may arise from weak accretion, shocks, or ionisation by evolved stellar populations, whereas a star-forming
classification does not exclude a weak or intermittent nucleus. 

Accordingly, the nuclear criterion is expressed as
\noindent
the Milky-Way-compatible nuclear-state class,
$\mathcal{N}_{\rm MW}$, while persistent Seyfert-like or quasar-like activity with a substantial Eddington ratio constitutes a failure of Stage III. This formulation distinguishes compatibility with the nuclear evolutionary regime of the Milky Way from a requirement for one specific present-day spectroscopic class:
\begin{enumerate}
\item[(i)] a quiescent nucleus;
\item[(ii)] a star-forming or composite nucleus without a dominant AGN;
\item[(iii)] LINER-like or radiatively inefficient nuclear activity;
\item[(iv)] an LLAGN consistent with a low AGN duty cycle.
\end{enumerate}

An initial simulation-based comparison illustrates the model dependence of jointly reproducing the adopted stellar-and black-hole-mass ranges. In the TNG50-1 snapshot at $z=0$, none of the 388 galaxies with $2\times10^{10}\leq M_\star/M_\odot\leq8\times10^{10}$ and an unambiguously identified single central black hole satisfies $10^{6}\leq M_{\rm BH}/M_\odot\leq10^{7}$; the lowest black-hole mass in this subsample is $1.34\times10^{7}\,M_\odot$. Applying the same mass cuts to the SIMBA $m100n1024$ snapshot identifies 52 validated central galaxies. The presence of qualifying systems in SIMBA shows that the adopted stellar- and black-hole-mass combination can be reproduced within a cosmological galaxy-formation model, even though it is absent from the selected TNG50 sample. So, the contrasting outcomes motivate examining the effects of black-hole seeding, accretion, feedback, and assembly histories, together with differences in simulation volume, resolution, and sample selection \citep{Nelson2019TNG50, Dave2019SIMBA}.

The combination of a massive stellar disc and a relatively low-mass SMBH in the Milky Way emphasises that galaxy growth and black-hole growth are related but non-equivalent processes. A low present-day SMBH mass need not imply substantial mass loss; it may reflect limited cumulative growth through accretion and black-hole mergers. Present-day nuclear quiescence and accumulated SMBH mass should therefore be treated as distinct diagnostics.
 
\textbf{Brief conclusion.} Nuclear activity provides an important constraint within the MWNeT methodology because the present-day state of a galactic nucleus represents only one phase of a potentially episodic SMBH activity cycle. Stage~III therefore excludes galaxies hosting persistently powerful AGNs while retaining quiescent and weakly active nuclei broadly compatible with the Milky Way. Evidence for past nuclear activity, AGN duty cycles, and fossil signatures of previous accretion events provides a deeper evolutionary diagnostic and is considered at Stage~VII. This distinction allows present-day nuclear similarity to be used as a practical selection criterion without assuming that genuine MWNeTs must reproduce the instantaneous activity state of Sgr~A$^\ast$.

\subsection{Global Spectrophotometric and Dynamical Constraints} 

The global properties of the Milky Way commonly used in MWA searches, including structural sizes, luminosities, colours, stellar mass, SFR, and characteristic circular velocity, are summarised in Table~\ref{tab:MW_properties}, while the corresponding selection ranges adopted for MWNeTs are given in Table~\ref{tab:MWNeTcriteria}. Among these quantities, stellar mass, star-formation rate, and characteristic circular velocity are particularly informative because they constrain the accumulated stellar content, present-day star-formation activity, and total gravitational potential of a galaxy, respectively.

\subsubsection{Global properties}
The use of global spectrophotometric and dynamical parameters in MWA searches is motivated by the assumption that galaxies with similar stellar masses, luminosities, colours, star-formation rates, structural properties, and characteristic rotation velocities occupy comparable regions of the general disc-galaxy population. These quantities are linked through fundamental scaling relations, including the Tully--Fisher relation, the stellar mass--size relation, the star-forming main sequence, and the baryonic Tully--Fisher relation \citep{Tully1977, Bell2003, Dutton2011, McGaugh2012}. Accordingly, classical MWA searches have employed different combinations of these observables. \cite{Mutch2011} used stellar mass and structural parameters, \cite{Licquia2015} selected MWAs using stellar mass and SFR, and \cite{Boardman2020a} combined stellar mass with bulge-to-disc ratio.

Imposing more constraints simultaneously generally yields smaller samples with closer present-day similarity to the Milky Way. \cite{Fraser2019} identified 176 MWAs among more than one million SDSS galaxies using stellar mass, morphology, and bulge-to-disc ratio, while \cite{Boardman2020a} found no MaNGA galaxies satisfying four simultaneous selection criteria. Despite these limitations, global spectrophotometric and dynamical parameters remain indispensable because they are available for large galaxy samples and efficiently reduce the parent population before applying more observationally demanding evolutionary diagnostics. Within the proposed hierarchy, Stage~IV assesses whether the galaxies retained after Stages~I--III reproduce the Milky Way's present-day global properties closely enough to qualify as MWNeT candidates. Their evolutionary interpretation is subsequently tested and refined through the complementary diagnostics of Stages~V--VII.

\citet{Tuntipong2024} developed a nearest-neighbour approach to rank SAMI galaxies according to their similarity to the Milky Way based on stellar mass, star formation rate, bulge-to-total ratio, and disc effective radius. They presented the ten closest galaxies as illustrative examples of the method; however, the approach itself can be applied to select any number of MWAs depending on the adopted similarity threshold. This ranking-based strategy provides a continuous similarity scale instead of a fixed binary classification and is therefore well suited to identifying progressively closer Milky Way counterparts.

The MWNeTs framework represents a further step in this development by incorporating evolutionary diagnostics into the similarity assessment, thereby identifying not only the closest structural analogues but also the nearest MW evolutionary counterparts.

\subsubsection{Stellar mass, star formation rate, and characteristic circular velocity}
Among the classical MWA parameters, stellar mass and star-formation rate deserve special attention because they are frequently derived through SED fitting. 

In our approach, the SED serves a dual role in MWNeT studies. First, SED fitting provides estimates of fundamental galaxy properties, including stellar mass, star-formation rate, dust content, and stellar-population characteristics. Second, the shape of the integrated SED is treated as a potential evolutionary diagnostic because it encodes the combined contributions of stellar populations, interstellar dust, gas, and nuclear activity. Similarity in SED shape therefore provides an additional constraint on the physical state and long-term energetic balance of candidate galaxies, although it cannot by itself establish similarity in evolutionary history (see in detail below).

The stellar mass of the Milky Way is currently estimated to be of the order of 
$(5$--$7)\times10^{10}\,M_{\odot}$, placing it among massive disc galaxies. Stellar mass is one of the most frequently used parameters in MWA searches because it reflects the integrated history of star formation and strongly correlates with numerous galaxy properties, including metallicity, luminosity, and halo mass. Consequently, stellar mass is commonly regarded as a primary constraint in MWA selection \citep{Licquia2015, Boardman2020a, Fraser2019}.

The present-day star-formation rate of the Milky Way is estimated at $1$--$3\,M_{\odot}\,{\rm yr}^{-1}$ \citep{Licquia2015}. Together with stellar mass, the SFR determines a galaxy's position on the star-forming main sequence and therefore characterises its current evolutionary state. Galaxies with substantially higher or lower SFRs may exhibit different gas-accretion histories, feedback processes, and evolutionary pathways. For this reason, SFR has become a standard parameter for MWA searches.

The characteristic circular velocity of the Milky Way,$V_{\rm c}\simeq220$--$240~{\rm km,s^{-1}}$, provides a direct measure of its total gravitational potential and dark-matter halo mass. Rotation velocity is central to scaling relations such as the Tully--Fisher and baryonic Tully--Fisher relations and therefore serves as an important dynamical constraint in MWA studies \citep{Tully1977, McGaugh2012, Licquia2016}. Recent Gaia-based measurements have further demonstrated that the detailed shape of the Galactic rotation curve contains valuable information about the distribution of baryonic and dark matter in the Milky Way. While we treat the rotation velocity as a classical global parameter, we consider rotation-curve morphology separately as an evolutionary diagnostic.

Spectroscopic observations complement global photometric measurements by constraining stellar populations, metallicity, star-formation activity, gas excitation, and ionisation mechanisms. Emission-line diagnostics based on $\mathrm{H}\alpha$, $\mathrm{H}\beta$, $[\mathrm{O\,III}]$, $[\mathrm{N\,II}]$, $[\mathrm{S\,II}]$, and $[\mathrm{O\,I}]$ allow galaxies to be classified as star-forming, composite, LINER-like, Seyfert, or weakly active systems using BPT-type diagnostic diagrams. Within the MWNeT methodology, such information contributes both to the present-day spectrophotometric characterisation of candidate galaxies and to the nuclear-activity constraints applied at Stage~III.

\subsection{Towards Milky Way Near Twins: Evolutionary and Multiwavelength Diagnostics}

Evolutionary diagnostics extend the classical MWA concept beyond present-day global parameters by probing the physical processes that have shaped a galaxy over cosmic time. In general, similar global properties do not imply similar evolutionary histories. Consequently, diagnostics sensitive to secular evolution, nuclear activity, chemo-dynamical evolution, and multiwavelength fossil signatures are essential for identifying  MWNeTs.

Table~\ref{tab:MWNeTcriteria} clearly illustrates the multidimensional nature of the MWNeT search problem. In preference to considering all potentially relevant parameters simultaneously, we arrange them hierarchically according to their physical significance, progressively identifying the principal characteristics of MWNeTs: \textit{isolated or weakly isolated barred spiral galaxies with absent or weak nuclear activity and small SMBHmass, and comparable global spectrophotometric, structural, and dynamical properties to those of the Milky Way}. This hierarchical strategy substantially reduces the dimensionality of the search problem.

But not only. For example, if Stages I, III, and V were omitted from this Table, and all remaining parameters were treated as a single multidimensional parameter space, the task would reduce to a conventional search for galaxies with the smallest offsets from the Milky Way by simultaneously comparing numerous global observables. In contrast, the parameters introduced in Stages~I and III, together with the evolutionary diagnostics discussed in this subsection, are physically motivated filters that constrain the evolutionary pathway of candidate galaxies instead of merely their present-day observable properties.

The advanced diagnostics discussed below span Stages~V--VII of the MWNeT hierarchy: panchromatic SED similarity, rotation-curve similarity, and evolutionary fingerprints and fossil tracers. These stages constitute the most flexible and evolving part of the framework. The application od these diagnostics remains challenging due to limited theoretical understanding of certain evolutionary processes and the heterogeneous quality and availability of observational data. 

Despite the unprecedented volume of astrophysical information accumulated by modern multiwavelength sky surveys, precisely identifying genuine MWNeTs remains challenging. However, once a class of galaxies satisfying the hierarchical criteria described in the previous stages has been identified, the search for additional evolutionary diagnostics becomes considerably more tractable. If specific evolutionary indicators are consistently found among several well-established MWNeTs, they may be regarded as common characteristics of the class and incorporated into the MWNeT framework as new evolutionary diagnostics. In other words, features identified in individual MWNeTs may subsequently be examined in other candidates to determine whether they represent common evolutionary characteristics of the class rather than properties of a single galaxy.

Nevertheless, recent observational and theoretical studies provide a growing set of diagnostics for identifying galaxies that share not only the Milky Way's present-day global properties but also important aspects of its evolutionary pathway.

Studies that reconstruct the Milky Way as an unresolved external galaxy have shown that integrated observations alone cannot fully recover key chemo-dynamical information. In particular, the Galactic disc chemical bimodality is not directly recovered from integrated-light IFU data; the underlying line-of-sight velocity distribution (LOSVD) is more complex than a standard Gauss--Hermite parametrisation; and higher-order kinematic moments together with several stellar-population properties remain difficult to constrain from integrated spectra \citep{Lian2023, Khoperskov2026}. These results imply that galaxies with similar global photometric properties, stellar masses, colours, or star-formation rates may nevertheless possess substantially different chemo-dynamical structures and evolutionary histories. This further motivates extending the classical MWA concept into the MWNeT framework by incorporating independent evolutionary diagnostics.

Several studies have attempted to incorporate evolutionary information into the search for Milky Way analogues. \citet{Pilyugin2023} introduced the concept of evolutionary Milky Way analogues (eMWAs), in which chemical abundance gradients were added to the classical structural and photometric selection criteria. \citet{Zhou2023} compared MWAs using their star-formation histories and chemical evolution inferred from MaNGA and APOGEE data, demonstrating that galaxies with similar present-day global properties may nevertheless have experienced substantially different evolutionary histories. \citet{Boardman2020a} investigated the stellar populations, gas properties, and kinematics of previously selected MWAs, providing important insights into their physical characteristics. However, these studies addressed individual evolutionary indicators in isolation and did not develop a comprehensive methodology that integrates independent evolutionary diagnostics into a hierarchical framework for identifying MWAs.

In our work, we consider several groups of advanced diagnostics, including chemo-dynamical properties, SED measurements and shapes, three-dimensional kinematics and rotation-curve phenomenology, multiwavelength signatures and peculiarities, and fossil tracers of galaxy evolution. Together, these  omplementary diagnostics comprise Stages~V--VII and test the evolutionary interpretation of the MWNeT candidates selected through Stages~I--IV.

\subsubsection{3D kinematics and Characteristic circular velocity } 
The Milky Way rotation curve exhibits several characteristic features that distinguish it from a simple universal profile. In the inner Galaxy, a pronounced excess above a smooth circular-velocity distribution is associated with the dynamical influence of the Galactic bar and bulge. \citet {Sofue2013} presented an earlier comprehensive analysis of the Galactic rotation curve and mass distribution from the central black hole to the outer Galaxy, while \citet{Sofue2020} later provided a review and unified analysis of the Milky Way rotation curve and dark-matter distribution.

At larger Galactocentric distances (up to $\sim25$ kpc), the rotation curve remains approximately flat, indicating the dominant contribution of the dark matter halo \citep{Eilers2019}. \citet{McGaugh2016} adopted this rotation-curve morphology as one criterion in the search for MWAs. In addition, local deviations and small-scale features reflect the influence of spiral arms, radial motions, and non-axisymmetric structures \citep{Sofue2021, Mroz2019}. Recent Gaia-based studies have further revealed subtle variations associated with the complex three-dimensional kinematics of the Galactic disc \citep{Fedorov2021, GaiaCollaboration2023, Fedorov2023, Dmytrenko2023, MrOz2024, Denyshchenko2024, Drimmel2025}.

In our recent Gaia-based study of red giants and subgiants near the Galactic plane \citep{Dmytrenko2025, Fedorov2025}, we derived the spatial variations of the centroid kinematic parameters, including the gradients $\partial V_R/\partial\theta$ and $\partial V_{\theta}/\partial\theta$. These measurements enable the reconstruction not only of the classical Galactic rotation curve, $V_{\rm rot}(R)$, but also of its azimuthal dependence, $V_{\rm rot}(R,\theta)$, thereby extending the traditional one-dimensional description to a two-dimensional representation of Galactic dynamics. The function $V_{\rm rot}(R,\theta)$ contains information on the dynamical influence of the Galactic bar, spiral structure, streaming motions, radial migration, and possible relics of past interactions. Consequently, the azimuthal variation of the rotation curve may provide a more sensitive evolutionary diagnostic than the classical radial curve alone. If future integral-field spectroscopy and high-resolution surveys provide comparable two-dimensional kinematic information for nearby MWNeTs, it could provide one of the strongest dynamical constraints on their evolutionary similarity to the Milky Way.

\textbf{Brief conclusion.} Rotation-curve morphology, particularly its azimuthal dependence, probes the dynamical history of MWNeTs rather than only their present-day mass distribution.

\subsubsection{Chemo-dynamical evolutionary parameters and peculiarities} 
Chemo-dynamical properties provide one of the most direct links between a galaxy's present-day structure and its evolutionary history. Unlike global photometric parameters, metallicity distributions, stellar-population gradients, kinematic substructures, and globular-cluster (GC) systems preserve long-lived signatures of gas accretion, radial migration, secular evolution, and merger events. Consequently, they represent some of the most physically meaningful diagnostics for identifying Milky Way Near Twins.

Several studies have shown the limitations of global scaling relations for establishing genuine Milky Way similarity. \citet{Hammer2007} suggested that the Milky Way is deficient in stellar mass, angular momentum, disc radius, and peripheral metallicity for its rotation velocity, whereas \citet{Licquia2016} found it to be unusually compact for its luminosity and rotation velocity. In contrast, \citet{McGaugh2016} showed that the Galaxy is broadly consistent with the standard Tully--Fisher and size--mass relations. These differing conclusions further motivate chemo-dynamical diagnostics that probe galaxies' formation and evolutionary history beyond their present-day positions in global scaling relations.

Galactic archaeology shows that the Milky Way's present-day structure reflects the combined effects of its assembly history, stellar migration, secular evolution, and chemical enrichment \citep{Minchev2016}. Reconstructing this history requires a chemodynamical approach that combines stellar kinematics, ages, and chemical abundances. This evolutionary perspective motivates including chemodynamical diagnostics, in addition to structural parameters, in MWNeT identification.

Among chemo-dynamical parameters, the metallicity provides particularly important constraints on evolution. While the Galactic centre exhibits one of the highest metallicities among nearby spiral galaxies \citep{Pilyugin2007}, the outer disc shows comparatively low oxygen abundance and a steeper metallicity gradient than most giant spirals \citep{Pilyugin2023, Pilyugin2024}. Using MaNGA galaxies, \citet{Pilyugin2023} identified several MWAs with similarly low peripheral metallicities, suggesting that the outer-disc chemical structure may represent one of the distinctive characteristics of the Milky Way.

Recent Gaia observations, together with cosmological simulations, have considerably refined our understanding of the Galaxy's chemo-dynamical evolution. They support a three-stage formation scenario consisting of an early turbulent protogalaxy, a kinematically hot old disc, and a younger dynamically cold disc, consistent with the Gaia--Sausage--Enceladus merger event and early disc formation \citep{Chandra2023, Semenov2024}. Additional analyses based on Gaia, APOGEE, H3 Survey, and TNG50 simulations further revealed significant differences among the stellar populations of the Galactic bar, bulge, inner disc, and halo, together with signatures of radial migration, metal-rich inner populations, and kinematically distinct halo components \citep{Queiroz2021, Wylie2022, Wylie2022a, Rix2024, Han2024}. These studies demonstrate that galaxies with similar global properties may nevertheless possess substantially different chemo-dynamical structures.

The Galactic globular-cluster system provides an additional fossil record of the Milky Way's assembly history. Approximately $150$--$180$ GCs populate the bulge, disc, and halo \citep{Harris1996}. Their ages, metallicities, and orbital properties preserve evidence for both the earliest stages of Galaxy formation and subsequent accretion events involving systems such as Gaia--Sausage--Enceladus, Sagittarius, Sequoia, and Kraken \citep{Minniti1995, MarinFranch2009, ForbesBridges2010, Barbuy2018, Massari2019, Kruijssen2020}. Recent $\phi$-GPU simulations further demonstrate that the orbital evolution and mass loss of Galactic GCs are strongly influenced by the time-dependent Galactic potential and assembly history \citep{Ishchenko2023a, Ishchenko2024}. 

\textbf{Brief conclusion.} Chemo-dynamical properties preserve the evolutionary memory of galaxies over timescales much longer than those accessible through global photometric parameters alone. Mean stellar metallicity, radial metallicity gradients, the structure of the Galactic bar and disc, and the multiphase circumgalactic medium. Long-lived GC systems as fossil tracers of galaxy evolution provide an additional diagnostic for identifying realistic MWNeTs.

\subsubsection{Multiwavelength Evolutionary Signatures}
Multiwavelength observations provide complementary constraints on the physical state and evolutionary history of Milky Way-like galaxies because different spectral domains trace different components of the baryonic matter, star-formation activity, dust content, magnetic fields, and nuclear processes. Within the MWNeT methodology, however, multiwavelength information is not used simply to increase the number of observables. Its principal role is to identify physically informative signatures that constrain whether candidate galaxies share the Milky Way's present-day energetic balance and evolutionary characteristics.

\textbf{Ultraviolet and optical} observations provide complementary constraints on recent star formation, stellar populations, morphology, colours, metallicity, and ionised-gas properties. In the MWNeT framework, these data primarily contribute to the global spectrophotometric characterisation of candidate galaxies and to reconstructing their stellar-population and chemo-dynamical properties. The Legacy Survey of Space and Time (LSST) at the Vera C. Rubin Observatory will substantially extend these opportunities by providing deep, homogeneous, multi-epoch optical imaging over a large fraction of the sky. Its combination of depth, spatial coverage, and time-domain information will enable systematic searches for low-surface-brightness tidal features, stellar streams, faint satellite systems, extended disc structures, and nuclear variability in nearby Milky Way-like galaxies. These observations will provide new constraints on merger histories, environmental interactions, secular evolution, and episodic nuclear activity, thereby extending optical observations from present-day spectrophotometric characterisation toward evolutionary diagnostics of MWNeT candidates.

\textbf{Radio continuum and high-energy observations} provide access to physical processes that are poorly constrained at shorter wavelengths. Radio emission traces cosmic-ray populations, magnetic fields, synchrotron halos, and signatures of large-scale outflows, whereas X-ray and gamma-ray observations probe hot gas, compact objects, weak nuclear activity, and energetic feedback processes. Such observations are especially important when searching for fossil structures and signatures of past energetic events, which we consider separately below as advanced evolutionary diagnostics.

\textbf{Infrared diagnostics} are particularly important because they trace dust heated by star formation and nuclear activity while remaining relatively insensitive to dust extinction. Owing to the large amount of dust in the Galactic plane, IR observations reveal the internal structure. Surveys and missions such as GLIMPSE \citep{Benjamin2005, Churchwell2009}, WISE \citep{Wright2010}, Herschel \citep{Molinari2011}, and \textit{Planck} \citep{Planck2011}, together with comprehensive reviews of Galactic structure \citep{Bland2016} and dust physics \citep{Draine2003, Draine2007}, have revealed the Galactic bar, inner ring, central molecular zone, large-scale filamentary structures, and the distribution of cold dust and star-forming regions. Because they operate in the near-IR, these observations are much less affected by dust extinction than optical surveys and provide a reliable representation of old stellar populations and the underlying stellar mass distribution. Consequently, IR surveys have become an indispensable tool, for example, recent studies based on SDSS and MaNGA have shown that galaxies with similar optical properties may exhibit substantially different IR characteristics \citep{Licquia2015mw, FraserMcKelvie2019, Boardman2020a, Zhou2023}. 

For our methodology, the Two Micron All Sky Survey (2MASS; \citealt{Skrutskie2006}) is particularly important because it is much less sensitive to dust extinction than optical observations. It provides an excellent tracer of old stellar populations and the underlying stellar mass distribution, allowing the identification of the Galactic bar and other internal structures of the Milky Way. Owing to these advantages, 2MASS has been extensively used in studies of the morphology and IR properties of MWAs \citep{FraserMcKelvie2019, Boardman2020a, GarmaOehmichen2023bar} as well as in our works for isolated galaxies \citep{Vavilova2009, Pulatova2015, Vavilova2016, Vasylenko2020, Pulatova2023, Kompaniiets2025, Kompaniiets2026}.

The advent of the \textit{James Webb Space Telescope} (JWST) has opened new opportunities for infrared studies of galaxies \citep{Gardner2006}. Its exceptional sensitivity and angular resolution in the near- and mid-IR domains enable detailed investigations of stellar populations, dust properties, obscured star formation, and the interstellar medium \citep{Pontoppidan2022, Rigby2023, Markov2023}. JWST observations may reveal previously inaccessible signatures of galaxy formation and evolutionary history \citep{Whitler2023, Leja2023, Lee2023} and, combined with multiwavelength surveys and SED modelling, provide valuable constraints for the detailed characterisation and validation of MWNeT candidates.

 Among the multiwavelength diagnostics considered here, the \textbf{integrated spectral energy distribution} provides a common framework for comparing the global energy distributions of the Milky Way and candidate MWNeTs. We therefore consider the SED separately as an integrated evolutionary fingerprint.

\subsubsection{Spectral energy distribution as an integrated energy fingerprint}
 
 \textbf{The spectral energy distribution} plays a dual role in the MWNeT methodology. SED fitting provides estimates of fundamental physical parameters, whereas the shape of the integrated SED is tested here as a higher-order evolutionary diagnostic and energetic fingerprint of a galaxy. Thus, SED-shape similarity may provide one of the most restrictive diagnostics in the transition from broad MWA populations toward genuine MWNeTs. However, constructing a reliable MW reference SED remains challenging because photometry is heterogeneous and observational constraints are uneven across the electromagnetic spectrum.
 
The Milky Way SED extends from the UV to the radio and includes contributions from young and old stellar populations, thermal dust emission, molecular and atomic gas, synchrotron radiation, and weak nuclear activity. Owing to the moderate star formation rate and low level of AGN activity, the Galactic SED is broadly consistent with that of a moderately star-forming barred spiral galaxy with specific mid-IR and X-ray emission \citep{Bland2016, Licquia2015mw, Natale2022}. Additionally, the Milky Way approximately follows the well-established far-infrared (FIR)--radio correlation observed in normal star-forming galaxies, which reflects the close connection among star formation, dust heating, cosmic rays, and magnetic fields \citep{Helou1985, Yun2001, Tabatabaei2017}. 

Modern SED modelling commonly relies on the energy-balance principle, $E_{\rm absorbed}\approx E_{\rm reradiated}$, implemented in widely used codes such as MAGPHYS \citep{daCunha2008}, CIGALE \citep{Boquien2019}, BAGPIPES \citep{Carnall2018}, and Prospector \citep{Johnson2021Prospector}. These tools enable physically motivated estimates of star-formation histories, stellar populations, dust properties, and AGN contributions and provide the modelling framework required for quantitative comparisons between the Milky Way and MWNeT candidates.

SED properties of MWAs were explored in several studies (e.g. \cite{Licquia2015, Fraser2019, Boardman2020b}) using selected spectral ranges. \cite{Fielder2021} constructed the Galaxy’s own SED by applying machine-learning techniques to a training sample of MWAs. However, the MWA training sample was selected based solely on stellar mass and SFR, which may introduce systematic biases into the reconstructed Milky Way SED and highlight the limitations of parameter-based analogue selection. Recent studies combining JWST data with BAGPIPES and Prospector have highlighted the importance of physically motivated SED modelling for interpreting stellar populations, dust attenuation, and star-formation histories \citep{Harvey2024, GimenezArteaga2022, Khoperskov2026}. 

The potential of SED-based diagnostics for identifying MWNeTs has been illustrated in our work \citep{Kompaniiets2026} for the nearby barred spiral galaxy NGC~3521. We extended the SED measurements beyond the conventional UV--cm-radio range considered in previous studies to include the metre and decametre radio domains.\footnote{We applied aperture photometry to GALEX, SDSS, WISE, Spitzer/MIPS, Herschel/PACS, SPIRE, and VLA images. To constrain the decametre emission and derive an upper limit in the 24--32 MHz band, we used observations obtained in January and February 2022 with the Ukrainian T-shaped Radio Telescope (UTR-2) \citep{Albergaria2025} and developed an additional radio prescription module in CIGALE.} The integrated SEDs of the Milky Way and NGC~3521 show remarkable overall similarity, with residual offsets in the far-UV and near the far-IR dust peak. This result favours the SED-shape comparison as a complementary evolutionary diagnostic and makes NGC~3521 a useful pathfinder for future MWNeT studies.

\textbf{Cosmological simulations} provide an additional perspective that increasingly incorporates radiative transfer and synthetic photometry to predict galaxy SEDs. Large numerical projects such as IllustrisTNG \citep{Nelson2019TNG50}, EAGLE\citep{Schaye2015}, TNG50 \citep{Pillepich2019}, FIRE-2 \citep{Hopkins2018}, and SIMBA \citep{Dave2019SIMBA}, combined with modern radiative-transfer techniques implemented in codes such as classical SUNRISE \citep{Jonsson2006a} and SKIRT \citep{Camps2015, Camps2020} make it possible to generate realistic multiwavelength SEDs and to compare them directly with observations. The field of cosmological simulations of Milky Way analogues is already well developed. However, more realistic modelling of Milky Way-like galaxies, together with the identification of observational counterparts selected using evolutionary diagnostics, may improve the connection between simulated and observed galaxy populations and help reduce galaxy-related systematic uncertainties relevant to precision cosmology, including those discussed in the context of current cosmological tensions \citep{H0DN, DiValentino2025}.

\textbf{Brief conclusion.} The integrated SED may serve as an energetic fingerprint of MWNeTs, complementing structural, dynamical, and chemo-dynamical diagnostics. The combination of observations, modern SED-fitting techniques, cosmological simulations, and radiative-transfer modelling may ultimately enable the construction of physically motivated ``virtual Milky Ways" and provide a bridge between theoretical galaxy-formation histories, external MWNeT candidates, and the Milky Way itself.

\subsubsection{Multiwavelength peculiarities as fossil tracers of the evolutionary history }
Beyond its global multiwavelength properties, the Milky Way exhibits several peculiar large-scale structures that may preserve information about past energetic events. These include radio spurs and loops, synchrotron structures, the North Polar Spur, and the Fermi and eROSITA bubbles, as well as signatures of past nuclear activity and possible TDE-driven accretion episodes. Such features can be regarded as fossil tracers of the Galaxy's evolutionary history.
 
\textbf{Radio spurs and synchrotron structures.} In the radio wave domain, synchrotron emission from relativistic cosmic-ray electrons becomes dominant, making diffuse and aged nonthermal structures particularly prominent. Because synchrotron spectra are steep, these structures often appear brighter and more extended at frequencies of a few tens of MHz than in GHz surveys. Consequently, low-frequency observations uniquely probe aged cosmic-ray populations, weak magnetic fields, and diffuse halo structures that may no longer be detectable at higher frequencies \citep{Braude1978Survey, Haslam1982, Iwashita2024}.

Large radio spurs and loops trace Galactic magnetic fields, cosmic-ray populations, halo circulation, superbubbles, and large-scale outflows. Low-frequency facilities such as UTR-2 \citep{Braude1978, Konovalenko2016} and LOFAR \citep{vanHaarlem2013} provide particularly valuable constraints on diffuse synchrotron structures associated with the Galactic halo. The North Polar Spur is one of the best-known examples and is observed in radio continuum, polarised emission, and X-rays. Its origin remains debated, with interpretations ranging from local superbubble activity to large-scale Galactic-centre outflows \citep{Miroshnichenko2009, Lallement2022, Iwashita2024}, and gaseous plumes associated with star-forming regions located at Galactocentric distances of $\sim3$--5 kpc, whose morphology is shaped by buoyancy and differential Galactic rotation \citep{Churazov2024}. Regardless of its precise origin, the North Polar Spur illustrates how large-scale multiwavelength structures may preserve information about energetic processes that are no longer directly observable.

\textbf{Fermi and eROSITA bubbles.} The discovery of the giant Fermi bubbles \citep{Su2010} and their X-ray counterparts, the eROSITA bubbles \citep{Predehl2020}, revealed clear evidence for powerful energy injection into the Galactic halo. Extending tens of kiloparsecs above and below the Galactic plane, these structures constitute some of the most remarkable manifestations of past activity in the Milky Way. Their morphology, energetics, and multiwavelength appearance indicate that the Galaxy has experienced episodic feedback processes on timescales of several million years (see the recent review by \citealt{Sarkar2024}).

The origin of the Fermi and eROSITA bubbles remains under debate. They are commonly interpreted as relics of past AGN-like activity associated with Sagittarius~A$^\ast$ or nuclear starburst episodes and Galactic winds \citep{BlandHawthorn2019, Sarkar2024}. TDE-driven episodes of enhanced SMBH accretion have also been proposed as possible contributors to large-scale Galactic outflows and bubble energetics \citep{Sazonov2012, Komossa2015, Alexander2017}. Their gamma-ray, X-ray, microwave, polarised-radio, and synchrotron counterparts reveal relativistic particles, magnetic fields, and hot plasma \citep{Su2010, Predehl2020, Sarkar2024}. If analogous structures are identified in candidate MWNeTs, they may provide powerful diagnostics of past nuclear activity and long-term feedback that present-day global galaxy properties cannot infer.

\textbf{Tidal disruption events and relic nuclear activity.}  Although no tidal disruption event (TDE) has been directly observed in the Milky Way, theoretical estimates suggest an occurrence rate of $\Gamma_{\rm TDE}\sim10^{-5}-10^{-4},\mathrm{yr}^{-1}$, corresponding to approximately one event every $10^{4}$--$10^{5}$ years \citep{Komossa2015, Alexander2017}. Several studies point out that the Galactic Centre experienced episodes of significantly enhanced activity in the past \citep{Baganoff2003, Ho2008, Yuan2014}. Relic manifestations of such activity may be preserved in radio, microwave, X-ray, and gamma-ray structures extending far beyond the Galactic plane \citep{BlandHawthorn2019}. Such events can temporarily enhance the accretion rate onto Sagittarius A$^\ast$ and generate energetic outflows capable of affecting the Galactic environment. In particular, TDEs have been proposed as one mechanism responsible for forming the Fermi bubbles and related large-scale structures in the Galactic halo \citep{Sazonov2012}.

More than 100 TDE candidates have been reported, and their census continues to expand with optical, X-ray, IR, and radio time-domain surveys \citep{Gezari2021, Hammerstein2023, Masterson2024, EylesFerris2025, GuillochonOpenTDECatalog}.  Our recent study of the isolated barred galaxy UGC~11487, which we consider a promising MWNeT candidate, illustrates the potential of this approach. Phase-resolved WISE/NEOWISE observations reveal a structured circumnuclear dust response to an obscured TDE-like accretion event, demonstrating that time-dependent infrared diagnostics can recover information on the geometry of the circumnuclear medium and recent accretion history of galaxies with no dominant persistent AGN component. Such diagnostics may therefore complement large-scale fossil tracers by probing more recent episodes of nuclear activity in candidate MWNeTs.

\textbf{Brief conclusion.} Multiwavelength peculiarities, including diffuse radio structures, the Fermi and eROSITA bubbles, and relic signatures of past nuclear activity or TDE-like events, may preserve evolutionary information inaccessible from present-day global parameters and integrated SEDs. By probing energetic processes on different spatial and temporal scales, these fossil tracers provide complementary constraints on the feedback history and activity cycles of the Milky Way and candidate MWNeTs.

\subsection{From Milky Way Near Twins Back to the Milky Way}
The proposed methodology is inherently iterative rather than unidirectional. Once we identify Milky Way Near Twins, they become astrophysical laboratories for refining our understanding of the Milky Way's evolutionary history. Thus, the search for MWNeTs does not terminate with the identification of suitable galaxies; rather, it initiates a reverse process in which selected near twins provide new constraints on the formation and evolution of our own Galaxy.

Since we observe the Milky Way from within, many of its global properties and evolutionary characteristics remain difficult to determine directly. Conversely, we observe external galaxies as complete systems, allowing us to measure their global structure and integrated properties directly. MWNeTs may therefore provide the external perspective required to place the Galaxy in a broader cosmological context and to test whether its observed properties and evolutionary history are typical or exceptional. The transition from MWAs to MWNeTs thus represents not only a search for increasingly realistic analogues, but also a way to improve our understanding of the Milky Way itself \citep{Bland2016, Vavilova2024}.

Comparative studies of MWNeTs may help refine estimates of the Galaxy's global parameters, including stellar and halo masses, star-formation rate, bulge-to-total ratio, and bar properties. More importantly, such systems may constrain aspects of the Milky Way's evolutionary history, including merger events, secular evolution, radial migration, and the formation history of the Galactic bar \citep{Helmi2020}. MWNeTs may also provide insights into the activity cycles of Sagittarius~A$^\ast$, the occurrence of tidal disruption events, and past AGN-like activity, as well as gas accretion, the hot circumgalactic medium, and the missing-baryon problem \citep{Tumlinson2017}. Our recent panchromatic forward-modelling study of NGC~3521 \citep{Kompaniiets2026} illustrates this approach by demonstrating how observations of a close MWNeT candidate can constrain global properties that are difficult to reconstruct directly for the Milky Way.

Particular importance may be attached to multiwavelength peculiarities and fossil tracers of energetic events, including radio spurs, synchrotron halos, and structures analogous to the Fermi and eROSITA bubbles. Their identification in external MWNeTs could provide independent constraints on the feedback history and nuclear-activity cycles of the Milky Way \citep{Sarkar2024}. More generally, comparative studies of MWNeTs address a fundamental question in Galactic astronomy: is the Milky Way a typical barred spiral galaxy, or does it occupy a special place among galaxies of similar mass and morphology? Recent studies suggest that the Milky Way may be an ``unusually typical'' galaxy \citep{Licquia2016}. In this sense, the search for Milky Way Near Twins is a natural extension of the Copernican principle and bridges Galactic astronomy, extragalactic studies, and cosmology.

Different evolutionary diagnostics preserve signatures of physical processes operating over different timescales and therefore probe complementary layers of a galaxy's evolutionary memory. Global photometric properties primarily characterise its present-day observable state, whereas rotation curves constrain its dynamical structure, chemo-dynamical properties record the formation and evolution of the disc and stellar populations, globular-cluster systems preserve evidence of past accretion events, the integrated SED reflects the long-term energetic balance, and multiwavelength fossil tracers retain signatures of previous episodes of nuclear activity and feedback. MWNeTs therefore provide a comparative system in which complementary evolutionary records observed in external galaxies can be used to reconstruct aspects of the Milky Way's history that are inaccessible from any single diagnostic.

Future observational facilities and surveys, including the Vera C. Rubin Observatory (LSST), the Nancy Grace Roman Space Telescope, the Square Kilometre Array (SKA), JWST, Athena, future Gaia-like astrometric missions, and integral-field spectroscopic surveys, will substantially expand the multiwavelength, kinematic, and chemo-dynamical information available for nearby galaxies. Combined with increasingly sophisticated analysis and artificial-intelligence techniques, these data will enable more detailed identification and characterisation of MWNeTs and strengthen their use as external laboratories for reconstructing the Milky Way's evolutionary history.

The MWNeT framework therefore closes the loop between Galactic and extragalactic astronomy. The Milky Way serves as the astrophysical reference for identifying its nearest evolutionary counterparts, while MWNeTs, in turn, provide the external observational perspective required to refine our understanding of the Galaxy's structure, formation history, and long-term evolution. Milky Way Near Twins serve as observational laboratories for investigating the evolutionary history of the only galaxy that we cannot observe from the outside: the Milky Way.

\section{Applicability and interpretation of the criteria}
\label{sec:framework_limitations}

The hierarchy distinguishes mandatory selection criteria from supplementary diagnostics. Mandatory criteria define the minimum compatibility required at a given stage: a securely established incompatibility may exclude a galaxy from the corresponding candidate set. Supplementary diagnostics strengthen or qualify the assessment, but a departure from a representative Milky Way value does not automatically imply rejection. The role of each criterion should be specified before assessing individual candidates and applied consistently across the sample. Stages~I--IV are mandatory for initial candidate selection, but not every diagnostic within these stages is a mandatory rejection criterion. The ranges in Table~\ref{tab:MWNeTcriteria} provide reference intervals for assessing compatibility; departures in supplementary diagnostics require qualification or automatic exclusion. Stages~V--VII strengthen the evolutionary assessment, while missing higher-stage measurements do not invalidate candidate status established through Stages~I--IV.

Missing measurements must be distinguished from physical incompatibility. An unmeasured mandatory property remains unconstrained, whereas an uncertain or definition-sensitive estimate requires qualification rather than automatic rejection. Supplementary diagnostics likewise contribute only where sufficiently reliable and comparable measurements are available. Globular-cluster systems, circumgalactic-medium properties, and other demanding evolutionary diagnostics are included in Stage~VII because they probe assembly and feedback histories, while comparable observations remain sparse for external galaxies. They therefore provide advanced tests of the evolutionary interpretation rather than prerequisites for initial candidate selection. Their absence limits the strength of the evolutionary assessment without invalidating compatibility established at earlier stages.

Paper~II applies these distinctions to NGC~3521, NGC~2903, UGC~11487, and CGCG~047-026, separating observational completeness from physical compatibility.

\section{Conclusions}

We introduce \textit{Milky Way Near Twins} (MWNeTs) as a physically motivated class of galaxies resembling the Milky Way in their present-day properties and exhibiting observational signatures consistent with broadly comparable evolutionary pathways. The principal conceptual advance is to treat present-day similarity as necessary but insufficient for identifying the Galaxy's closest evolutionary counterparts.

The proposed seven-stage hierarchy combines candidate selection with evolutionary assessment. Stages~I--IV are mandatory for initial selection and assess environmental, morphological, structural, nuclear, and global spectrophotometric and dynamical compatibility. Stages~V--VII test and refine the evolutionary interpretation through panchromatic SED similarity, rotation-curve similarity, and evolutionary fingerprints and fossil tracers. Mandatory criteria are distinguished from supplementary diagnostics, and missing observations are distinguished from physical incompatibility.

We introduce \textit{evolutionary memory} to describe complementary observational records of processes operating on different timescales. The integrated SED characterises the present-day energy distribution and provides model-dependent constraints on the star-formation history; rotation curves constrain dynamical structure; chemo-dynamical signatures and globular-cluster systems record aspects of disc formation and accretion; and circumgalactic gas and multiwavelength fossil tracers constrain gas cycling, feedback, and past nuclear activity. Together, these records define a galaxy's \textit{evolutionary fingerprint}. Their convergence strengthens the hypothesis of evolutionary similarity without uniquely reconstructing a common history.

The framework is extensible and iterative. New diagnostics may be incorporated as their physical interpretation and observational accessibility improve. The Milky Way provides the reference system, while candidate near twins offer external laboratories for testing aspects of Galactic structure and evolution that are difficult to investigate from our internal viewpoint. Paper~II presents the observational application to
NGC~3521, NGC~2903, UGC~11487, and CGCG~047-026.

The search for MWNeTs extends the long-standing scientific quest to understand whether our cosmic environment is unique. Just as the discovery of exoplanets motivated searches for Earth-like planets and planetary systems resembling the Solar System, the identification of MWNeTs motivates the search for galaxies resembling our own in both their present-day properties and their evolutionary histories. By linking Galactic astronomy, extragalactic astronomy, and cosmology, the MWNeT framework advances this question from whether galaxies like the Milky Way exist to whether they reached their present state through broadly similar evolutionary pathways.

\begin{acknowledgements}
This study was supported by the National Research Foundation of Ukraine (Project No. 2023.03/0188). Iryna Vavilova thanks Prof. Leonid S. Pilyugin (MAO NASU, Ukraine) for inspiring discussions in 2021 on the use of metallicity gradients as potential evolutionary diagnostics of MWAs. She is also grateful to Prof. Johan H. Knapen and Prof. John E. Beckman (IAC, Spain) for stimulating discussions during a seminar in 2025 on the scientific motivation for searching for the closest Milky Way counterparts, when the MWNeT concept was still taking shape. The authors are grateful to Dr. Alessandro B. Romeo, Dr. Sujeeporn Tuntipong, and Prof. Yaroslav S. Yatskiv for their insightful comments and constructive suggestions on an earlier version of this manuscript.

This research has made extensive use of the SAO/NASA Astrophysics Data System.
\end{acknowledgements}

\bibliographystyle{aa} 
\bibliography{MWNeTs} 

\appendix
\clearpage
\onecolumn

\section{Reference properties of the Milky Way}
\label{app:MWprop}

\setlength{\LTpre}{0pt}
\setlength{\LTpost}{0pt}

\begin{longtable}{@{}p{0.28\textwidth}p{0.68\textwidth}@{}}
\caption{Reference properties and evolutionary characteristics of the Milky Way relevant to the identification of Milky Way Near Twins.}
\label{tab:MW_properties}\\
\hline
Property & Typical value or characteristic \\
\hline
\endfirsthead

\multicolumn{2}{c}{\tablename~\thetable\ -- continued}\\
\hline
Property & Typical value or characteristic \\
\hline
\endhead

\hline
\multicolumn{2}{r}{\footnotesize Continued on the next page}\\
\endfoot

\hline
\endlastfoot

Morphological type & Barred spiral galaxy, approximately SAB(rs)bc--SBbc ($T\approx4$); the exact external morphology is model-dependent \citep{deVaucouleurs1978,Bland2016,Shen2020} \\
Spiral structure & Tracer-dependent spiral pattern: four major gaseous and young-stellar arms are commonly identified, whereas the old stellar disc may show a more prominent two-armed component; the Local Arm forms an additional segment \citep{Bland2016, Xu2016, Shen2020} \\
Absolute visual magnitude & $M_V\sim-20.5$ to $-21.5$; representative integrated luminosity estimates remain model-dependent because the Milky Way is observed from within \citep{deVaucouleurs1978, Licquia2015mw, Bland2016} \\
Total stellar luminosity & $L_\star\sim(1$--$3)\times10^{10}\,L_\odot$; model-dependent integrated estimate \citep{deVaucouleurs1978,Licquia2015mw,Bland2016} \\
Colour index & Integrated optical colour, approximately $(B-V)_T\sim0.5$--$0.7$; SDSS-based estimates give ${}^{0}(g-r)\simeq0.68$ \citep{deVaucouleurs1978,Licquia2015mw} \\
Disc scale length & Exponential stellar-disc scale length $R_{\rm d}\sim2.5$--$3.5\,{\rm kpc}$, depending on the tracer and adopted disc model \citep{Bland2016,Licquia2015mw} \\
Stellar disc radius & Main stellar-disc radius $\sim15$--$20\,{\rm kpc}$; stars and gas can be traced beyond this radius depending on the tracer and surface-brightness limit \citep{Bland2016, Shen2020} \\
Effective radius & Half-light radius $R_{\rm eff}\sim4$--$6\,{\rm kpc}$; the value depends on the adopted surface-brightness model and treatment of the inner Galaxy \citep{deVaucouleurs1978, Bland2016} \\
Isophotal radius & $R_{25}\approx11$--$12\,{\rm kpc}$; mainly a historical extragalactic-style estimate and not equivalent to the full stellar-disc radius \citep{deVaucouleurs1978, Bland2016} \\
Bar morphology & Boxy/peanut-shaped bulge associated with the Galactic bar; the Milky Way is generally interpreted as hosting a barred inner structure rather than a massive classical bulge \citep{Wegg2015, Bland2016, Sanders2019} \\
Bar semi-major axis & $a_{\rm bar}\sim4$--$5.5\,{\rm kpc}$; estimates depend on the tracer population and on whether the bar length is defined photometrically or dynamically \citep{Wegg2015, Portail2017, Sanders2019, Lucey2023} \\
Bar pattern speed & $\Omega_{\rm bar}\sim33$--$40\,{\rm km\,s^{-1}\,kpc^{-1}}$; recent kinematic measurements favour a relatively low pattern speed, although the inferred value remains method- and model-dependent \citep{Sanders2019, Clarke2022, Lucchini2024} \\
Thin-disc scale height & $h_{z,\rm thin}\sim0.2$--$0.4\,{\rm kpc}$; tracer- and population-dependent \citep{Bland2016,Mackereth2017} \\
Thick-disc scale height & $h_{z,\rm thick}\sim0.8$--$1.3\,{\rm kpc}$; tracer- and population-dependent \citep{Bland2016,Mackereth2017} \\
Bulge-to-total ratio & $B/T\sim0.1$--$0.2$; dominated by the bar/boxy-peanut structure, with a small model-dependent classical-bulge contribution \citep{Shen2010, Bland2016} \\
Characteristic circular velocity & $V_{\rm c}(R_0)\sim220$--$240\,{\rm km\,s^{-1}}$ near the Solar Galactocentric radius; recent Gaia DR3 Cepheid measurements give $V_{\rm c}(R_0)=236.8\pm0.8\,{\rm km\,s^{-1}}$, although the inferred value remains dependent on the tracer population and Galactic mass modelling \citep{Eilers2019, Mroz2019, Fedorov2025, Feng2026} \\
Rotation-curve morphology & Approximately flat to mildly declining over much of the Galactic disc, with tracer-dependent substructure and evidence for a stronger decline at large Galactocentric radii; the outer rotation curve remains sensitive to disequilibrium, non-axisymmetry, and modelling assumptions \citep{Jiao2023, Ou2024, Koop2024, Fedorov2025, Feng2026} \\
Orbital period at the Solar radius & $T_{\rm orb}(R_0)\sim220$--$250\,{\rm Myr}$, estimated from the adopted Solar Galactocentric radius and circular velocity \citep{Bland2016,Eilers2019} \\
Stellar mass & $M_\star\sim(5$--$7)\times10^{10}\,M_\odot$; representative estimates depend on the adopted IMF, stellar-population modelling, and Galactic structural model \citep{Licquia2015mw, Bland2016} \\
Virial halo mass & $M_{200}\sim(0.8$--$1.5)\times10^{12}\,M_\odot$; estimates remain method- and model-dependent \citep{Bland2016,Callingham2019,Wang2020} \\
Star-formation rate & $\mathrm{SFR}\sim1$--$3\,M_\odot\,\mathrm{yr}^{-1}$ \citep{Chomiuk2011,Licquia2015mw,Bland2016} \\
Neutral atomic hydrogen mass & $M_{\rm HI}\sim(5$--$8)\times10^{9}\,M_\odot$; estimates depend on the adopted Galactic H\,{\sc i} distribution and integration radius \citep{Kalberla2009,Bland2016} \\
Stellar metallicity & Population- and position-dependent; broadly near-solar, with $[\mathrm{Fe/H}]\sim-0.3$ to $+0.3$ in the thin disc at the Solar radius, systematic radial and vertical gradients, and distinct chemo-dynamical populations \citep{Bland2016, Hayden2015, Queiroz2020} \\
Gas-phase metallicity gradient & Negative radial oxygen-abundance gradient across the Galactic disc, with representative estimates of approximately $-0.04$ to $-0.06\,{\rm dex\,kpc^{-1}}$, depending on the tracer and radial range \citep{Pilyugin2014, Esteban2018, ArellanoCordova2020} \\
Cold-gas fraction & $f_{\rm gas}=M_{\rm cold\,gas}/(M_\star+M_{\rm cold\,gas})\sim0.1$--$0.2$; the value depends on the adopted H\,{\sc i} and H$_2$ masses, helium correction, and integration radius \citep{Bland2016} \\
Nuclear activity type & Nearly quiescent SMBH or extremely low-luminosity active nucleus, often discussed in the LLAGN context (Sgr~A$^\ast$) \citep{Baganoff2003, Ho2008, Yuan2014} \\
Supermassive black-hole mass & $M_{\rm SMBH}\simeq4.0\times10^{6}\,M_\odot$ \citep{Ghez2008,Genzel2010,GravityCollaboration2019,Becerra-Vergara2021} \\
Accretion rate & $\dot{M}\sim10^{-9}$--$10^{-7}\,M_\odot\,\mathrm{yr}^{-1}$; the inferred value depends on the spatial scale and accretion-flow model \citep{Baganoff2003, Yuan2014} \\
Eddington ratio & $\lambda_{\rm Edd}\ll10^{-6}$; Sgr~A$^\ast$ is among the most sub-Eddington SMBHs known in nearby galactic nuclei \citep{Ho2008, Yuan2014} \\
Environment & Member of the Local Group; no ongoing major interaction with a nearby comparable-mass galaxy, although weak interactions with the Magellanic Clouds and other satellites occur \citep{McConnachie2012} \\
Cosmic-web location & Located in the Local Sheet, adjacent to the Local Void and embedded in the surrounding filamentary environment \citep{Tully2008, Tully2014, Pomarede2020} \\
Satellite system & LMC, SMC, and more than 60 known dwarf satellite galaxies; the census remains observationally incomplete \citep{McConnachie2012, Simon2019, DrlicaWagner2020} \\
Globular-cluster system & Approximately $150$--$180$ known globular clusters, including in-situ and accreted subpopulations that preserve evidence of the Milky Way's assembly history \citep{Harris1996, Kruijssen2019, Massari2019} \\
Circumgalactic medium & Extended multiphase CGM containing hot, warm, and cool gas, with evidence for gas accretion, Galactic outflows, and baryon cycling between the disc and halo \citep{Tumlinson2017, Bland2016} \\
Multiwavelength fossil tracers & Large-scale structures and signatures of past energetic activity, including radio spurs and loops, the North Polar Spur, Fermi and eROSITA bubbles, X-ray echoes, and Galactic-centre outflow structures, which preserve evidence of episodic nuclear activity and feedback \citep{Ponti2019, Predehl2020, Lallement2022, Iwashita2024, Sarkar2024} \\
\end{longtable}

\end{document}